\documentclass[12pt,preprint]{aastex}

\include{psfig}

\shorttitle{A star-forming filament at $z=1.8$}
\shortauthors{Stevens et al.}

\begin{document}

\title{A filamentary structure of massive star-forming galaxies \\ 
associated
with an X-ray absorbed QSO at z=1.8}

\author{J. A. Stevens}
\affil{Astronomy Technology Centre, Royal Observatory,
Blackford Hill, Edinburgh EH9 3HJ, UK}
\email{jas@roe.ac.uk}
\author{M. J. Page}
\affil{Mullard Space Science Laboratory, University College
London, Holmbury St. Mary, Dorking, Surrey RH5 6NT, UK}
\email{mjp@mssl.ucl.ac.uk}
\author{R. J. Ivison}
\affil{Astronomy Technology Centre, Royal Observatory,
Blackford Hill, Edinburgh EH9 3HJ, UK}
\email{rji@roe.ac.uk}
\author{Ian Smail}
\affil{Institute for Computational Cosmology, University of Durham, South Road, Durham
DH1 3LE}
\email{ian.smail@durham.ac.uk}
\and
\author{F. J. Carrera}
\affil{Instituto de Fisica de Cantabria (Consejo Superior de
Investigaciones Cientificas-Universidad de Cantabria), Avenida de los Castros
39005 Santander, Spain}
\email{carreraf@ifca.unican.es}

\begin{abstract}
The genesis of spheroids is central to our understanding of galaxy formation --
they are relatively simple systems, containing about half the stellar mass of
the Universe. A major subset of spheroids, massive elliptical galaxies, are
preferentially found in clusters where they exhibit old, coeval stellar
populations suggesting that they formed synchronously at early epochs.  Here we
report SCUBA submillimeter imaging of a region around a $z=1.8$ X-ray selected
QSO. The image reveals a remarkable $\sim400$~kiloparsec long chain of galaxies
each with an obscured star-formation rate sufficiently high to build a massive
spheroid in less than 1~Gyr.  The large over-density of these galaxies relative
to expectations for a random field implies they probably reside in a structure
associated with the QSO. We suggest that this star formation is associated with
galaxy mergers or encounters within the filament, such as those predicted by
the popular hierarchical model of galaxy formation. Our observations suggest
that strong absorption in the X-ray spectra of QSOs at
high-redshifts may result from a veil of gas thrown up by a merger or
merger-induced activity, rather than an orientation-dependent obscuring
torus. It is argued that these systems are the precursors of elliptical
galaxies found today in the core regions of all rich galaxy clusters.
\end{abstract}


\section{Introduction}

Hierarchical structure formation models are predicated on the assumption that
structures grow via the gravitational collapse of small inhomogeneities in a
Gaussian random field.  Within such a model it is expected that galaxies and
larger scale structures will form at local maxima of the density field
\citep{kai84}. Inhomogeneities in such over-dense regions, that will go on to
become the most massive rich clusters at the present day, tend to collapse
earlier than similar inhomogeneities in less-dense regions \citep{kau96}.
Moreover, although the power spectrum of the fluctuations in these models are
scale free, detailed numerical simulations show that they naturally create
large-scale filamentary structures with the most massive bound systems, such as
rich clusters, forming at intersections and junctions of these filaments
\citep{col99}.  There is expected to be considerable cross-talk between
structures forming on different scales in these models: with galaxy halos
growing through major mergers within the larger-scale filaments and streaming
down these into the forming cluster.  These galaxies are most naturally
identified as the progenitors of the homogeneous population of old,
luminous elliptical galaxies which reside in rich clusters at the present day.
These models thus predict that the forming elliptical galaxies should be
distributed in a highly anisotropic fashion around the collapsing protocluster
\citep{wes94}.

To test these predictions requires observations in a waveband sensitive to
massive star formation at high redshifts.  Recent progress in submillimeter
astronomy has shown that star formation in massive galaxies at high redshifts
is not a luminous phenomenon at optical wavelengths because of the obscuring
effect of dust \citep{sma02}. The reprocessed starlight peaks at far-infrared
wavelengths in the rest frame of the source and is shifted into the
submillimeter waveband at $z > 1$. We also require a method of identifying an
over-dense region of the early Universe. One method is to target powerful
active galactic nuclei (AGN). Submillimeter images of the $\sim1$\ Mpc-scale
fields around such AGN (high-redshift radio galaxies -- HzRGs) at $z>3$ show
extended dust emission associated with the AGN, and luminous submillimeter
companions \citep{ivi00,ste03}, several of which are confirmed to be at the
same redshifts as the radio galaxies \citep{sma03,smi03}.  While these results
provide support for models of biased galaxy formation, they relate only to a
very rare class of tracer galaxies, powerful HzRGs at $z > 3$. Therefore, if we
are to investigate the formation of structures the size of typical clusters
then less extreme, lower mass signpost AGN at lower redshifts must be
identified and their environments mapped at submillimeter wavelengths. With
these points in mind we have initiated a programme to image the fields around
$1.5 < z < 3$ QSOs detected in the submillimeter waveband \citep{pag01}. Here
we report the first submillimeter imaging observations of one of these QSOs,
RXJ094144.51+385434.8 at $z=1.819$.

A Hubble constant $H_0=70$\ km\,s$^{-1}$\,Mpc$^{-1}$ and density
parameters $\Omega_{\Lambda}=0.7$ and $\Omega_{\rm m}=0.3$ are assumed
throughout this Letter.

\section{Observations and data reduction}

\subsection{Submillimeter imaging}

Observations were made with SCUBA \citep{hol99} at the JCMT in 2003 February and
March.  During all three nights on which data were gathered, weather
conditions were in the top quartile of those experienced on Mauna Kea. We used
`jiggle-map' mode to make simultaneous maps at 450 and 850~$\mu$m. The
secondary mirror was chopped $45''$ in right
ascension, and the antenna was nodded every 16 seconds. The resulting maps thus
show the negative off-positions of real sources (they have not been
`cleaned'). The atmospheric opacity was monitored with regular skydips, and
flux density calibration was made on each night with jiggle-map observations of
the standard source CRL618. 

Data were analysed with the {\sc starlink} package, {\sc surf}. Data were first
corrected for beam-switching, flat-fielded and extinction corrected. We then
removed residual sky emission and clipped each bolometer at the 5-$\sigma$
level before resampling the data onto a RA/Dec grid and despiking at the
3-$\sigma$ level. Pointing drifts were corrected for with linear interpolation
between the measured offsets (in all cases less than $2''$). After flux density
calibration, the data were rebinned to make the final signal and noise
maps. For this process we used an adapted version of the {\sc surf} task, {\sc
rebin} -- the standard deviation of the signal for each bolometer in each
observation was first calculated, the data were then rebinned onto a $1''$ grid
in RA/Dec coordinates, and the standard deviations were used to create weighted
signal and weighted noise maps. These maps were subsequently smoothed with a
$6''$ (at 850~$\mu$m) or $4''$ (at 450~$\mu$m) Gaussian.

\subsection{Optical and near-infrared imaging}

An $R$-band image of a $16'\times 16'$ region centered on the QSO was taken at
the 4.2-m William Herschel Telescope (WHT) using the Prime Focus Imaging Camera
on the night of 2003 May 28.  The total integration time was
3.7\,ks, the seeing was 1.0$''$ and the conditions were photometric.  The data
were reduced using standard {\sc iraf} scripts and calibrated using Landolt
faint standards \citep{lan92}.  The 5-$\sigma$ limiting magnitude for a
point-source in our $2.5''$ aperture is $R=25.4$.

A $K$-band exposure of the central $90''\times 90''$ part of the field was
obtained in photometric conditions with the UFTI imager on
the United Kingdom Infra-Red Telescope (UKIRT) on 2003 May 24.  The total
exposure time was 7.7\,ks in 0.6$''$ seeing yielding a 5-$\sigma$ limiting
magnitude of $K=20.5$.  The data were reduced using the {\sc orac-dr} pipeline
and the resulting mosaics combined with {\sc starlink ccdpack} tasks.  The data
were calibrated with observations of the UKIRT faint standard FS\,127
\citep{haw01}.

Galaxies were identified on the $K$-band frame using {\sc SExtractor} 
\citep{ber96} and total magnitudes estimated from {\sc best\_mag}.  The
positions were then used to measure colors within $2.5''$ diameter apertures
from the seeing-matched $R$/$K$ frames.

\section{Results and Discussion}

The 450- and 850-$\mu$m images reveal a remarkable structure of submillimeter
sources in the field of the QSO, particularly striking at 450\,$\mu$m
(Fig. 1). For the high redshift ($z > 3$) objects targeted to date
\citep{ste03}, the slightly different negative K-corrections that act on the
850- and 450-$\mu$m flux densities have resulted in poor quality 450-$\mu$m
images. This is because, while the 850-$\mu$m datum is shifted along the steep
Rayleigh-Jeans tail of the dust emission, effectively canceling out the
cosmological dimming, the 450-$\mu$m datum is shifted close to the peak of this
emission giving a marked fall-off of the flux density with redshift. In this
respect, the lower redshift of RXJ094144.51+385434.8 has worked to our
advantage, providing us with an image of the submillimeter dust emission with
unprecedented resolution ($8.5''$, equivalent to $\sim$70~kpc at
$z=1.8$).

We identify six sources with peak signal-noise greater than 3 in the 450-$\mu$m
image; they trace a `chain' of submillimeter galaxies with an extent of at
least 400~kpc (if at $z=1.8$).  Source names, coordinates and physical
quantities calculated from the images are presented in Table~1.  At 450~$\mu$m
the sources are well enough separated to allow good estimation of their
individual flux densities; this is not the case at 850\,$\mu$m where we give
combined flux densities for Nos 1,2 (the QSO) and 4,5. Quoted errors do not
include the calibration uncertainties which are about 10\% at 850~$\mu$m and
15--20\% at 450~$\mu$m.  Dust masses are calculated from the 450~$\mu$m flux
densities in the standard manner \citep{hil83}, adopting a value for the dust
mass absorption coefficient $\kappa_{100\mu m}=5.5$\,m$^2$\,kg$^{-1}$
\citep{dra84}. Far-infrared luminosities ($L_{\rm FIR}$) are calculated by
scaling the 450~$\mu$m emission with that of Mrk~231 which has $L_{\rm
FIR}=2.0\times10^{12}\ L_{\odot}$ (calculated from a grey-body fit to the
far-infrared--millimeter SED - the fitted dust temperature is 42~K). Using
Arp220 as a template reduces the $L_{\rm FIR}$ estimates by a factor $\sim1.6$.
The star-formation rates are calculated from SFR(M$_{\odot}$/yr)$=L_{\rm
FIR}/(5.8\times10^9L_{\odot}$) \citep{ken98}.

Each galaxy in this filament is itself a very massive ultraluminous infrared
galaxy (ULIRG) producing stars at a rate sufficient to build a massive spheroid
in $<1$ Gyr. Let us assess the significance of this structure. The density of
450-$\mu$m sources in this field is 1.4$\pm$0.6~arcmin$^{-2}$, disregarding the
QSO which was already known to be a submillimeter source.  The corresponding
density of sources found in `blank field' surveys with 450-$\mu$m flux
densities in excess of 30 mJy is uncertain, but best estimates lie in the range
$0.03-0.14$~arcmin$^{-2}$ \citep{sma02}. These figures thus imply an
order of magnitude over-density of luminous star-forming galaxies in the
$3.5$~arcmin$^{-2}$ field around RXJ094144.51+385434.8. We can investigate the
statistical significance of this over-density by calculating the combined
probability that these sources each exist at a distance, $d$, from the QSO
given the expectation from the known surface density of blank field
objects. The high end of the range given above, i.e. $n=0.14$~arcmin$^{-2}$, gives
the most pessimistic estimate of this probability (given by $P=1-\exp\{-\pi n
d^2\}$ for each source).  We find $P=2.9\times10^{-7}$ that the structure we
observe is a chance superposition of field objects ($P=6.5\times10^{-5}$ if
source 2 is excluded). This result provides very strong evidence that the
companion galaxies lie at the same redshift and in the same structure as the
QSO. Assuming this to be true, the star formation rate density calculated in a
500 kpc cube centered on this structure is $>1000$~M$_{\odot}$\ yr$^{-1}$\
Mpc$^{-3}$, about 3 to 4 orders of magnitude higher than found in unbiased
optical, infrared and submillimeter surveys \citep{ivi02,cha03}.  This field
thus provides a striking demonstration of the filamentary distribution of
forming galaxies in a high-density region at high redshift.
 
Can we find further support for the hierarchical model in our data?  In the
high resolution 450-$\mu$m image, the central QSO and at least one of the
spatially distinct submillimeter sources have complex morphologies, indicative
of either merging or interacting systems. These characteristics are analogous
to those observed in local ULIRGs \citep{jos90} although these have more
compact millimeter/submillimeter emission \citep{dow98} than observed here at
higher redshift. Moreover, these mergers must be occurring between similarly
massive (Table~1) and gas rich systems, indicating that this activity is
arising from major, rather than minor, mergers.  Images of the
RXJ094144.51+385434.8 field in the $R$- and $K$-band are shown in Fig.~2. The
submillimeter galaxies RXJ094144SMM3, RXJ094144SMM4/5 and RXJ094144SMM6 all
have counterparts identified as extremely red objects (EROs) \citep{els88} --
defined as having $(R - K) > 5.3$ \citep{poz00}. These objects are often found as
counterparts to submillimeter galaxies discovered in blank field surveys
\citep{sma99,ivi02}. The counterpart of RXJ094144SMM6 is particularly
interesting; it has the appearance of an advanced merger and is a blue/red
composite source, again similar to many counterparts of blank-field
submillimeter galaxies \citep{ivi02}. Similar evidence for a merger origin of
submillimeter galaxies has been seen in {\em Chandra\/} observations which show
them to be coincident with two or three X-ray sources -- interpreted as
obscured AGN buried in merging star-forming galaxies \citep{sma03,ale04}.  If
the submillimeter-detected galaxies are proto-ellipticals then at $z\sim2$ they
should contain growing black holes which shine as AGN
\citep{kau00,pag01}. However, the {\em ROSAT\/} data for this field show that
the QSO is the only luminous X-ray source amongst them. If the other dusty
galaxies contain AGN then they have to be of lower luminosity or more highly
obscured or both. Observations with the new generation of X-ray telescopes will
allow us to probe the evolutionary state of the black holes in these systems
\citep{sma03}.

These images are the first to be made of a high redshift, X-ray absorbed QSO at
submillimeter wavelengths. The importance of these objects relative to their
better studied X-ray unabsorbed (or optically selected) counterparts is only
just becoming clear. For matched ($1 < z < 3$) samples selected close to the
break of the X-ray luminosity function, the latter are about one order of
magnitude less luminous at submillimeter wavelengths. The most likely
interpretation is that the X-ray absorption in these objects is not linked with
orientation dependent obscuration - the `unified scheme' \citep{ant93} - but
rather with the evolutionary state of the galaxy \citep{pag04}. Our
submillimeter images indicate that this is indeed the case for
RXJ094144.51+385434.8 -- its submillimeter luminosity appears to originate in
large scale star formation associated with a major galaxy merger.

Finally, let us consider the fate of this structure. Deep {\it HST\/} images of
the fields around QSOs \citep{mcl01} suggest that their typical environments are
galaxy clusters of Abell richness class 0 \citep{abe58}. At the present day
such clusters contain within their virial radius of order 10 galaxies with
luminosity greater than or equal to L$^*$ \citep{chr03}. If the 6 submillimeter
luminous galaxies in the protocluster around RXJ094144.51+385434.8 continue to
form stars at the rates inferred from their submillimeter luminosities
(Table~1) then their transformation into L$^*$ ellipticals will be complete
within in a fraction of a Gyr. We can thus conclude that the coeval stellar
populations of the luminous cluster galaxies will be in place by redshift
1.7. Their subsequent passive luminosity evolution will naturally result in a
population of ellipticals with properties that match the core regions of
today's rich galaxy clusters.

\acknowledgments

The James Clerk Maxwell Telescope is operated by the Joint Astronomy Centre in
Hilo, Hawaii on behalf of the parent organizations PPARC in the United Kingdom,
the National Research Council of Canada and the Netherlands Organization for
Scientific Research. Near-infrared data were collected as part of the UKIRT
Service Programme.  UKIRT is operated by the Joint Astronomy Centre on behalf
of PPARC. The WHT is operated on the island of La Palma by the Isaac Newton
Group in the Spanish Observatorio del Roque de los Muchachos of the Instituto
de Astrofisica de Canarias. J.A.S. acknowledges support from PPARC, I.R.S. from
the Royal Society and F.J.C. from the Spanish Ministerio de Cienca y
Technolog\'{i}a, under project AYA2000-1690.

\clearpage

\begin{deluxetable}{ccccccccc}
\tabletypesize{\tiny}
\tablecaption{Source names, coordinates, submillimeter flux densities and parameters.}
\tablewidth{0pt}
\tablehead{
\colhead{No.\tablenotemark{a}} & \colhead{Source name\tablenotemark{b}} & \colhead{RA (J2000.0)\tablenotemark{c}} & \colhead{Dec (J2000.0)\tablenotemark{c}} & 
\colhead{$S_{850} (\rm mJy)$}& \colhead{$S_{450} (\rm mJy)$} & 
\colhead{Dust Mass ($M_{\odot}$)} & \colhead{$L_{\rm FIR}(L_{\odot})$}& 
\colhead{SFR$({\rm M}_{\odot}/{\rm yr})$}
} 
\startdata
1& RXJ094144SMM1& $09\ 41\ 44.96$ & $+38\ 54\ 39.0$ & & $44.3\pm7.6$ &$7.1\times10^8$& $1.6\times10^{13}$&$2700$\\
&&&&$13.4\pm1.5$\tablenotemark{d}&&&&\\
2& RXJ094144SMM2& $09\ 41\ 44.49$ & $+38\ 54\ 42.0$ & & $29.8\pm7.9$& $4.8\times10^8$& $1.1\times10^{13}$&$1800$\\
&&&&&&&&\\
3& RXJ094144SMM3& $09\ 41\ 44.00$ & $+38\ 54\ 24.8$ & $6.6\pm1.5$ &
$32.6\pm8.3$& $5.3\times10^8$& $1.1\times10^{13}$&$2000$\\ 
&&&&&&&&\\
4& RXJ094144SMM4& $09\ 41\ 46.63$ & $+38\ 54\ 15.8$ & & $33.3\pm8.9$& $5.4\times10^8$& $1.2\times10^{13}$&$2000$\\
&&&&$8.0\pm1.3$&&&&\\ 
5& RXJ094144SMM5& $09\ 41\ 46.16$ & $+38\ 54\ 16.0$ & & $37.8\pm9.4$& $6.1\times10^8$& $1.3\times10^{13}$&$2300$\\
&&&&&&&&\\
6& RXJ094144SMM6& $09\ 41\ 46.48$ & $+38\ 53\ 58.8$ & $6.5\pm1.8$ &
$39.7\pm8.6$& $6.4\times10^8$& $1.4\times10^{13}$&$2400$\\
&&&&&&&&\\
\enddata

\tablenotetext{a}{Source Nos are those printed on Fig.~1.}
\tablenotetext{b}{Source names are based on the {\em ROSAT\/} right ascension
position of the QSO.}  
\tablenotetext{c}{Source coordinates are taken from the
450-$\mu$m map which is centred on the QSO at $09^{\rm h}41^{\rm m}44^{\rm
s}.51$, $+38^{\circ}54'34''.8$ (J2000.0). The formal uncertainties on these
positions, given by FWHM/(S/N$\times2$) are only $0.7-1.2''$ although
it should be assumed that systematic effects will at least double these
values.}  
\tablenotetext{d}{The QSO - cf.\ $10.1\pm1.7$\ mJy from \citet{pag01}}
\end{deluxetable}

\clearpage

\begin{figure*}
\plotone{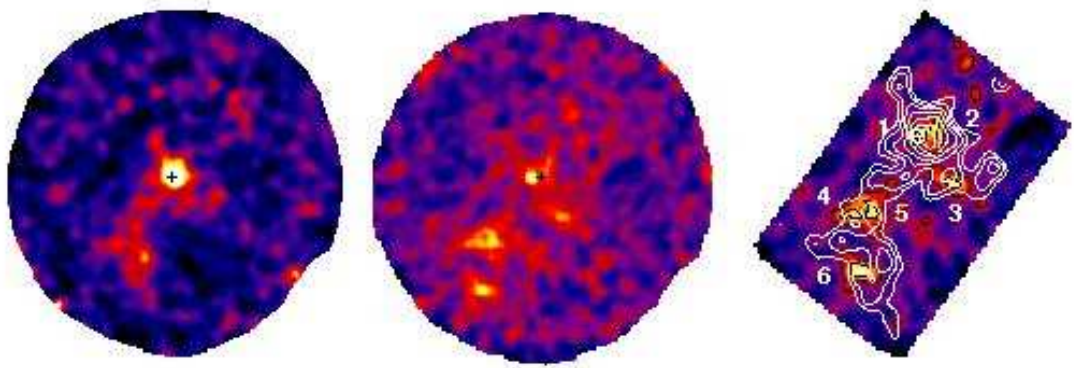}
\caption{Submillimeter imaging of the field around RXJ094144.51+385434.8. The
left-hand panel shows the SCUBA 850-$\mu$m image (diameter $\sim 150''$,
resolution $14.8''$). The middle panel shows the corresponding 450-$\mu$m image
(diameter $\sim 120''$, resolution $8.5''$). The right-hand panel shows a
signal-to-noise image ($1.5'\times1.0'$) at 450\,$\mu$m (greyscale with black
contours at 2, 3, 4 and 5$\sigma$) overlayed with the 850-$\mu$m
signal-to-noise contours at 2, 3, 4, 5, 6, 7 and 8$\sigma$. The extent of the
contoured region is $\sim400$~kpc. Numbers on the right-hand panel identify
450~$\mu$m sources with peak signal-to-noise greater than 3 - they correspond
to those listed in Table~1. The optical position of the QSO is marked with a
cross on the left-hand and middle panels.\label{fig1}}
\end{figure*}

\clearpage 

\begin{figure}
\vspace{-1.5in}
\setlength{\unitlength}{1in}
\begin{picture}(3.0,8.0)
\includegraphics{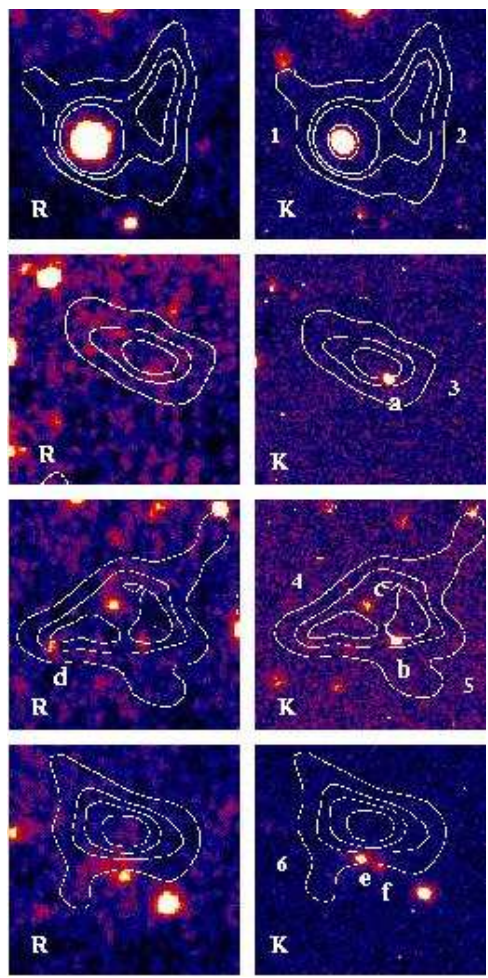}
\end{picture}
\caption[dum]{Optical and near-infrared imaging of the submillimeter galaxies
in the vicinity of RXJ094144.51+385434.8. The RA/Dec images are $22''$ square,
equivalent to 200~kpc at $z=1.8$.  Optical $R$-band images ($1''$ seeing) from
the WHT and near-infrared $K$-band images ($0.6''$ seeing) from UKIRT overlayed with
the 450\,$\mu$m contours from Fig.~1 (the 450\,$\mu$m image has been offset by
[$-4.9''$,$+0.3''$] to center the dust peak on the optical/near-infrared
source). Numbers on the panels are those from Fig.~1 and Table~1.  The
magnitudes of these counterparts are (a) $K=19.3$, $(R-K)=5.8$ (b) $K=19.9$,
$(R-K)=5.3$ (c) $K=20.3$, $(R-K)=3.9$ (d) $R=24.9$, $(R-K)<3.6$ (e) $K=18.8$,
$(R-K)=5.4$ (f) $K=19.5$, $(R-K)=4.6$.}
\label{fig2}
\end{figure}

\end{document}